\documentclass[12pt,twoside]{report}
\usepackage{svcon2e}
   \usepackage{amsthm,amssymb}
\usepackage{amsbsy,amsfonts} 
\usepackage{epsfig}
\usepackage[frenchb]{babel} \NoAutoSpaceBeforeFDP
\usepackage{alltt}
\usepackage{oldgerm}

\newcommand{\ca}[1]{{#1}}  
 
\newcommand{\cc}[1]{{#1}}

\def\cal{\mathcal}

\catcode`@=12
\makeatletter
\@addtoreset{equation}{section}

\catcode`@=12
\makeatletter
\def\@crosshairs{\vbox to 0pt{}}

\newcommand{\beq}{\begin{equation}}  
\newcommand{\eeq}{\end{equation}}
\newcommand{\bea}{\begin{eqnarray*}} 
\newcommand{\eea}{\end{eqnarray*}}
\newcommand{\bean}{\begin{eqnarray}}   
\newcommand{\eean}{\end{eqnarray}}

\makeatother

\newcommand\Ack{\medskip\noindent{\bf Acknowledgements}\enspace}
\newcommand\Problem[1]{\smallskip \noindent {\bf  Problem: }{\sl {#1}}}

\def\tr{{\rm tr}\,}

\def\rrangle{\rangle\!\rangle}

\def\mathph#1{{\tt math-ph/#1}}
\def\hepth#1{{\tt hep-th/#1}}
\def\qalg#1{{\tt q-alg/#1}}
\def\mathOA#1{{\tt math.OA/#1}}
\def\mathQA#1{{\tt math.QA/#1}}
\def\mathCT#1{{\tt math.CT/#1}}
                                       
\def\bi{\bar{i}}\def\bj{\bar{j}}
\def\bL{\bar{L}}\def\bCV{\overline{\CV}}
\def\bW{\overline W}

\def\CA{{\cal A}}\def\CD{{\cal D}}
\def\CE{{\cal E}}
\def\CH{{\cal H}}
\def\CI{{\cal I}}\def\CN{{\cal N}}\def\CR{{\cal R}}
\def\CU{{\cal U}}\def\CV{{\cal V}}
\def\Bbb#1{\mathbb{#1}}
\def\Ga{\alpha} \def\Gb{\beta}\def\Gc{\gamma}  
\def\za{\alpha} \def\zb{\beta} \def\zg{\gamma}

\def\gA{{\mathfrak A}}\def\gg{{\mathfrak g}}
\def\gAe{\gA^{{\rm ext}}}

\def\Fo{{}^{(1)}\!F}\def\Ft{{}^{(2)}\!F}
\def\tFo{{}^{(1)}\!\widetilde{F}}\def\tF{\widetilde{F}}

\def\tq{\tilde q}\def\btq{\tilde{q}}
\def\tn{\tilde n}
\def\tN{\widetilde{N}}\def\tV{\widetilde{V}}

\def\Fo{{}^{(1)}\!F}\def\Ft{{}^{(2)}\!F}
\def\tFo{{}^{(1)}\!\widetilde{F}}\def\tF{\widetilde{F}}

\def\tq{\tilde q}\def\btq{\tilde{q}}
\def\tn{\tilde n}
\def\tN{\widetilde{N}}\def\tV{\widetilde{V}}

\def\hN{{\hat N}}

\long\def\omit#1{{}}

\begin{document}
\pagenumbering{arabic}
  \thispagestyle{empty}
   \chapter{Conformal Field Theories, \hfil\break Graphs and Quantum Algebras}
\chapterauthors{
  \cc{ Valentina Petkova and 
Jean-Bernard Zuber }}

{\renewcommand{\thefootnote}{\fnsymbol{footnote}}

\footnotetext{\kern-19pt{\bf AMS Subject classification:} Primary 81R10,
81T40, 16W30, 20G42, 81R50}

\begin{abstract}

This article reviews some recent progress
 in our understanding of the structure of Rational Conformal
Field Theories, based on ideas that originate for a large part in the 
work of A. Ocneanu. The consistency conditions that generalize modular 
invariance for a given RCFT in the presence of various types of boundary
conditions --open, twisted-- are encoded in a system of integer
multiplicities that form matrix representations of fusion-like algebras.
These multiplicities are also the combinatorial data that enable one to
construct an abstract ``quantum'' algebra, whose $6j$- and $3j$-symbols
contain essential information on the Operator Product Algebra of the RCFT 
and are part of a cell system, subject to pentagonal identities. It 
looks quite plausible that the  classification of a wide class of 
RCFT amounts to a classification of ``Weak $C^*$- Hopf algebras''.

\end{abstract}


\section{Introduction}

For the last fifteen years or so, the study of Conformal Field 
Theory has been an amazingly active area of mathematical physics. 
Through all its connections with mathematics --infinite dimensional
algebras, quantum deformations of algebras, 
integrable systems, combinatorial identities
etc, and with its many applications in various fields of physics, 
from statistical mechanics to field and string theory, 
it offers a vivid evidence that exactly solvable physical 
problems  may be an extraordinary source of enrichment and 
cross-fertilization, which is also what Barry McCoy has been so
remarkably illustrating through his long and beautiful series of
works\dots

The present article is devoted to a presentation in physical terms 
of some algebraic structures 
which are quite suitable for the description of 
rational conformal field theories (RCFT's) 
with or without boundaries, and  which enable one to unravel 
 their hidden symmetries and to derive new information. 
The main idea may be summarised as follows.  
 Several consistency conditions encountered in RCFT lead to ``nimreps'',
an acronym standing for {\it non-negative integer valued matrix
representations}, of certain fusion algebras (and their extensions), 
and these nimreps may themselves be encoded
in graphs. These graphs are in the simplest case
Dynkin diagrams or their generalization; but more complicated patterns
that we call Ocneanu graphs also appear. 
The graphs provide
the combinatorial data needed for the construction of
an  algebra with two associative products,
the Ocneanu Double Triangle Algebra (DTA), interpreted also
as a pair of ``weak $C^*$-Hopf algebras" in duality \cite{bsz}.
There is a growing 
evidence that the data needed to define and characterize a RCFT are 
made of a ``cell system'' attached to these graphs
and subject to pentagonal identities.
These cells determine  the corresponding Ocneanu algebra, which
 can be interpreted as the quantum symmetry of the RCFT.

Most of the results of this review have been published in 
other papers \cite{pztw, pzcell},  
in which the reader will find more details. 
The present work was  inspired and deeply influenced 
 by  some recent work of A. Ocneanu
\cite{ocneanu:1998,ocneanu:2000}. Ocneanu's work originates in the 
study of topological invariants of 3-manifolds 
and unravels the common 
features of their construction with problems encountered in RCFT. Recently 
the reverse path was followed in
\cite{fffs:2000} to reformulate problems of
RCFT in terms of the associated 3d Topological Field Theory. 
The precise connection between these  approaches remains to be fully
clarified. 

\section{Chiral {R}CFT and its data}

Specification of a conformal theory begins with a certain number of 
chiral data. A {\it chiral algebra} \cc{$\gA\supseteq \hbox{Vir}$}
``containing'' the Virasoro algebra is given. It may be 
Vir itself, 
or some current (untwisted affine) algebra, or some 
$W$-algebra, or more generally a Vertex Operator Algebra. 
Beside the Virasoro generators $L_n$, the additional generators of $\gA$
 are denoted $W_n$. 
The assumption of {\it rationality} means that only a finite set $\CI$ of
irreducible representations (irreps) $\CV_i$ has to be considered
at a given value of the central charge. For each of these irreps, we 
introduce its {\it character} \cc{$\chi_i=\tr_{\CV_i} q^{L_0 -c/24}$},
$c$ the Virasoro central charge.
In a RCFT, these characters form a unitary representation of the 
modular group. If  $q=e^{2\pi i\tau}$, they transform 
by  a   symmetric unitary matrix $S$ under $\tau\to -1/\tau$: 
\cc{$\chi_i(\tau)=\sum_j S_{ij} \chi_j(-1/\tau)$}, and by a 
diagonal unitary matrix under $\tau\to\tau+1$.

A fundamental notion is that of {\it fusion of the representations} 
$\CV_i \star \CV_j$ and of the fusion coefficients which arise from 
 its decomposition into irreps {$ \CV_i \star \CV_j =\oplus_k
\CN_{ij}{}^k \otimes \CV_k$}.
Here $\CN_{ij}{}^k$ denotes  a finite dimensional multiplicity space; 
 by a small abuse  of notations, we shall henceforth write this kind 
 of decomposition as $ \CV_i \star \CV_j =\oplus_k N_{ij}{}^k
  \CV_k$, with $N_{ij}{}^k$  the integer dimension 
 of the space $\CN_{ij}{}^k$ .

These integers give the structure constants of an associative commutative
algebra which can be realized by the matrices $(N_i)_j{}^k$.
It has an identity $N_1=I$ corresponding to the ``vacuum representation''
$i=1$, i.e. the one whose lowest eigenvalue of $L_0$  is $0$.  
 The structure constants are
invariant under an involution map $i \to i^*$ s.t. $N_{ji}{}^1=\delta_{j
i^*}$ and are assumed to be given 
by the {\it Verlinde formula} $ N_{ij}{}^k =\sum_{l\in\CI} 
{S_{il}S_{jl}S_{kl}^*\over S_{1l}} 
$ 
\cite{verlinde}. 
 In the following, we shall make a repeated use of
a rephrasing of this formula: 
for a fixed $l\in\CI$, the ratios $S_{il}/S_{1l}$ form a 
one-dimensional representation of the fusion algebra

\beq
{{S_{il}\over S_{1l}} {S_{jl}\over S_{1l}}=\sum_{k\in\CI}
N_{ij}{}^k{S_{kl}\over S_{1l}}\ .}
\label{e:verl}
\eeq

\noindent
\begin{minipage}[t]{0.75\textwidth}
These fusion coefficients are also interpreted as giving 
the dimensions of the spaces of 
chiral vertex operators (CVO's) 
\cc{$\ \phi_{ij;t}^k(z) \ : \ \CV_j\mapsto  \CV_k\quad $} 
$\ t=1,\cdots,N_{ij}{}^k$ : 
\end{minipage} \hfill
\begin{minipage}[t]{0.20\textwidth}\vspace{0pt}
\centering \includegraphics[width=0.70\textwidth]{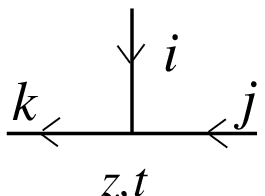}
\end{minipage} 

\medskip
\noindent
The braiding and fusing of these CVO's involve other matrices \\
\hbox{\raise -7mm\hbox{\epsfxsize=55mm\epsfbox{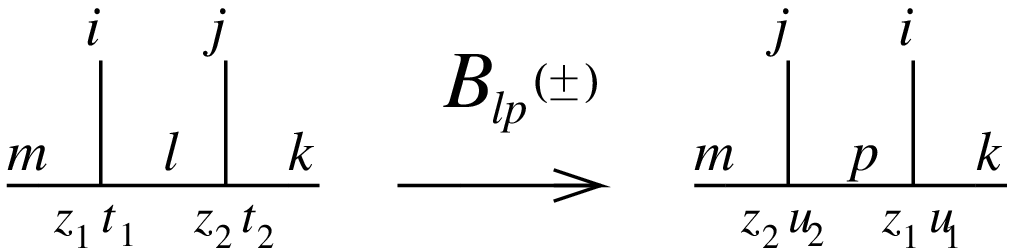}}} \qquad
\qquad
\hbox{\raise -7mm\hbox{\epsfxsize=55mm\epsfbox{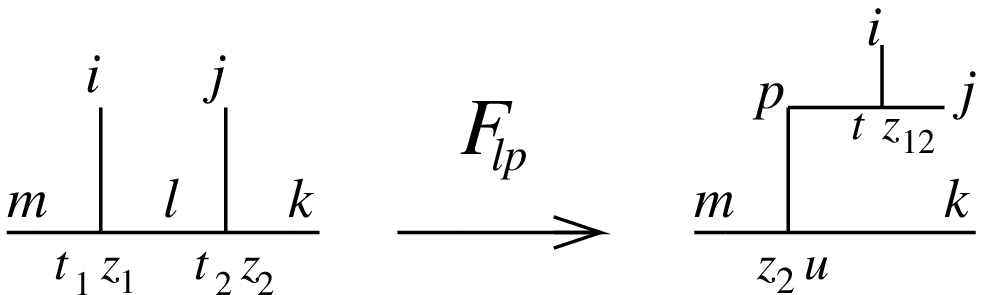}}}

\medskip
\noindent
{\bf Example:} Take \cc{${\gA}=\widehat{\gg}_k$},  the current algebra
based on the simple algebra $\gg$;
 $\CI$ labels  its set of integrable weights 
at level $k$; the $S$ matrix is given in \cite{kacpet}; 
the $F$ matrix turns out to be made of  the $6j$- symbols of the 
\cc{$\CU_q({\gg})$ quantum algebra for $q=\exp 2i\pi/(h+k)$}, with 
$h$ the dual Coxeter number.
In particular for $\widehat{sl}(2)_k$, \cc{$\CI=\{1,2,\cdots,k+1\}$},
and the $k$-dependent $S$-matrix is
\cc{$S_{ij}=\sqrt{2\over k+2} \sin {\pi ij\over k+2}$}.


\section{RCFT on a plane or a torus}

On a manifold without boundaries, like a plane 
or an infinite cylinder, or a torus, there are
\ca{\it two copies} of the chiral algebra $\gA$ at work, one relative to
the holomorphic (``left'') coordinates, the other to the
antiholomorphic ``right'' ones. It is traditional to affect the 
right copy of the algebra and of its representations, characters etc
with a bar. The Hilbert space of the theory is thus assumed to decompose 
as
\beq 
\CH=\oplus_{j,\bj\in \CI}\ Z_{j\bj}\ \CV_j\otimes \bCV_{\bj}
\label{e:hilb}
\eeq
with a finite dimensional multiplicity space $ Z_{j\bj}$ for each 
pair $(j,\bj)$.
  
\subsection{RCFT on a torus}
On a torus $\Bbb{T}=\Bbb{C}/(\Bbb{Z}+\tau \Bbb{Z})$, obtained by 
identifying the two ends of a finite segment of cylinder, and 
described by its modular 
 ratio $\tau$ or nome $q=e^{2i\pi \tau}$, 
a natural definition of the  
partition function of the system reads
\beq
Z=\tr_\CH q^{L_0-c/24} 
{q}^{*{\bL}_0-c/24}=\sum_{j\bj}
 Z_{j\bj}\ \chi_j(q)\ 
\big({\chi_{\bj}(q)}\big)^* 
\label{e:toruspf}
\eeq
As noticed by \cite{cardy:1986}, this partition function
must be \ca{modular invariant}. Moreover, the unicity of vacuum
imposes that $ Z_{11}=1$.

We are thus led to a well posed (but difficult!) problem \dots

\Problem{Classify the Modular Invariants}
$Z_{j\bj}\in 
\Bbb{Z}_{\ge 0}\,,\,
\quad Z_{11}=1 \  .$

As the representation of the modular group is unitary,
$Z_{j\bj}=\delta_{j\bj}$ is always a solution, the ``diagonal modular
invariant'',  in which all representations of $\CI$ appear once
in a left-right symmetric way: the corresponding ``diagonal theory''
has thus only spinless primary fields. 
The problem is to find all the other solutions. 
Unfortunately in spite of a constant flow of new results 
due  to the tenacity of T. Gannon \cite{gannon:19xx}, 
this approach has its limitations. Modular invariance is just a 
necessary condition, but is not sufficient 
to fully specify the theory, and 
other constraints may rule out some candidate solutions.
It is thus appropriate to look for additional restrictions.

For later reference, we recall that modular invariants come in two
types. Type I are block-diagonal invariants, of the form
$Z=\sum_a |\sum_i b_a^i\chi_i  |^2$, and are interpreted as the diagonal
invariants for some larger chiral algebra $\gAe \supset \gA$
whose characters decompose with branching coefficients $b_a^i
\in \Bbb{Z}_{\ge 0}$. Type 
II invariants are obtained from some invariant of type I, its ``parent'', 
through a twist of the right components with respect to the left by 
some automorphism $\zeta$ of the fusion rules of $\gAe$: 
$Z=\sum_a \sum_{i\bi} b_a^i b_{\zeta(a)}^{\bi}\chi_i \chi_{\bi}^*\, $.
(See the third reference in \cite{BE:xx} for examples beyond this
simple classification.)

\bigskip

\subsection{ Torus again, but with \cc{defect lines}} 

Let us now imagine that on the cylinder we introduce non-local operators
attached to non-contractible loops. 
These operators $X$ may be called defect or twist lines,
in the context of lattice models they have received the name of
 ``seams'', since they modify the nature of the boundary conditions 
as we close the cylinder into a torus \cite{pztw}.

\noindent
\begin{minipage}[t]{0.65\textwidth}
We demand that these operators, which act in the Hilbert space 
$\CH$ of equation 
(\ref{e:hilb}), commute with the action of $\gA\otimes\bar \gA$

\begin{eqnarray}
[L_n,X] & = & [\bar L_n,X]=0
\label{e:seam}\\
 \left[W_n,X\right] & = & [\bW_n,X]=0 \ .
\nonumber 
\end{eqnarray}

The interpretation of the former condition 
is that the action of $X$ is invariant under infinitesimal
\end{minipage} \hfill
\begin{minipage}[t]{0.30\textwidth}\vspace{0pt}
\centering \includegraphics[width=0.9\textwidth]{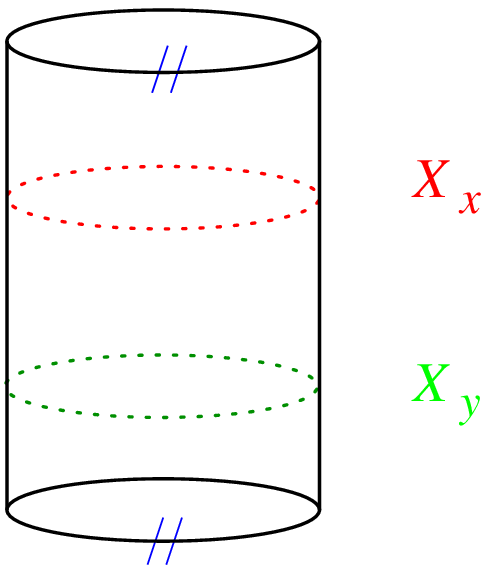}
\end{minipage} 
\vskip2mm \noindent
deformation of the contour, and the latter looks like a natural
extension to all the generators $W_n, \bW_n$ of the full 
chiral algebra. 

\medskip
Now, Schur's lemma leads to a complete characterization of these \cc{$X$}.
Their most general form is a linear combination of
$P^{(j,\bj;\Ga,\Ga')}$ which act as projectors 
between equivalent copies of \cc{$\CV_j\otimes\bar\CV_{\bj}$}
$$(\CV_j\otimes\bar\CV_{\bj})^{\Ga'}\mapsto(\CV_j\otimes\bar\CV_{\bj})^{\Ga},
\qquad\Ga,\Ga'=1,\cdots,Z_{j\bj}\ . $$
The number of independent $X$ is thus \cc{$ \sum Z_{j\bj}^2 =\tr Z
  Z^T$ }. We write a basis of such $X$ in the form 
\beq 
X_x= \sum_{j\bj,\Ga,\Ga'} {\Psi_x^{(j,\bj;\Ga,\Ga')}\over\sqrt{S_{1j}
S_{1\bj}}}\,  P^{(j,\bj;\Ga,\Ga')}
\label{e:PtoX}\,,
\eeq
where $\Psi$ is an invertible square matrix
of dimension $
\sum Z_{j\bj}^2$ and the denominator $\sqrt{SS}$ is 
introduced for later convenience.  Among these operators stands as 
a trivial solution the identity operator, which we label by $x=1$
and for which $\Psi_1^{(j,\bj;\Ga,\Ga')}=\delta_{\za\za'}\sqrt{S_{1j}
S_{1\bj}}$.

We then write the partition function on the torus with the insertion
of two defect lines in two alternative ways.   On the one hand
\beq
Z_{x|y}:=\tr_\CH\, X_x^\dagger X_y \ \tq^{L_0-c/24} {
  \tq}^{*\, {\bL}_0-c/24}\,, 
\eeq
where $\tq=\exp{-2\pi i\over \tau}$,
 is expressed through (\ref{e:PtoX}) as
\beq
Z_{x|y}
= \sum_{ j,\bj\in \CI\atop \za,\za'=1,\cdots,
Z_{j\bj}}
{\Psi_x^{(j,\bj;\za,\za')\, *}
\Psi_y^{(j,\bj;\za,\za')}\over S_{1j}S_{1\bj}} \
 \chi_j(\tq)\,  (\chi_{\bj}(\btq))^*\ .
\label{e:ZPsi}
\eeq
On the other hand, the system is also described by a Hilbert space
$${ \CH_{x|y}=\oplus_{i,\bi\in\CI}\, \tV_{i\bi^*;\, x}{}^y\, \CV_i\otimes
\bCV_{\bi}\ ,}$$
with some multiplicities $\tV_{ij;x}{}^y\in \Bbb{Z}_{\ge 0}\,,$
and in particular if the defects are trivial, 
\beq
\tV_{i\bi^*;\, 1}{}^1=Z_{i\bi}\ ,\label{e:tVZ} 
\eeq
the modular invariant matrix. Then the same partition function  reads
\beq 
Z_{x|y}=\tr_{\CH_{x|y}}\, q^{L_0-c/24}\,
 q^{*\, \bL_0-c/24}\,  =
\sum_{i,\bi\in\CI} \tV_{i\bi;\, x}{}^y \ \chi_i(q)\, (\chi_{\bi^*}(q))^*\ .
\label{e:ZtV}
\eeq

Note that the summation in (\ref{e:ZPsi}) runs over the set 
$\{(j,\bj;\Ga,\Ga')$, $\Ga,\Ga'=1,\cdots,Z_{j\bj}\}$
describing for $\alpha=\alpha'$ the physical spectrum  (\ref{e:hilb})
with its multiplicities, while in (\ref{e:ZtV})
it runs over all pairs of irreps of $\CI$.
Comparing the two expressions of $Z_{x|y}$ and assuming the
sesquilinear combinations $\chi_i\chi_{\bi}^*$ to be independent 
(which is justified only after a slight generalization of this
argument involving non-specialized characters \cite{BPPZ:2000,pztw}), we find 
 the consistency condition
\beq{  \tV_{i\bi;\, x}{}^y=
\sum_{j,\bj,\za,\za'}\,
{S_{ij}\,S_{\bi\bj} \over S_{1j}S_{1\bj}}\
\Psi_x^{(j,\bj; \za,\za')}\ \Psi_y^{(j,\bj;\za,\za')\, *}
\,,\qquad i,\bi\in\CI \,.
}
\label{e:consV}
\eeq
For $x=1=y$ this relation boils down to the modular invariance
property  $Z=S \, Z\, S^*$ of the torus partition function.
To proceed, we make the additional assumption that the
matrix $\Psi$ in the change of basis
of solutions of (\ref{e:PtoX}) $P^{(j,\bj;\za,\za')}\to X_x$ is unitary. 
Then (\ref{e:consV}) is the diagonalization formula of $\tV_{i\bi}$, 
whose eigenvalue ${S_{ij}S_{\bi\bj} / S_{1j}S_{1\bj}}$ has
multiplicity $Z_{j\bj}^2$.
We now recall (\ref{e:verl}) and conclude that the $\tV$ matrices 
form a ``nimrep'' of the double fusion algebra
\beq
 \tV_{i_1j_1} \tV_{i_2j_2}=
\sum_{i_3,j_3} N_{i_1i_2}{}^{i_3} N_{j_1j_2}{}^{j_3}\
\tV_{i_3j_3} \ .
\label{e:nimrepV}
 \eeq
The study of RCFT in the presence of defect lines has thus led us
to a new 
\vskip2mm 
\Problem{Classify the nimreps  \ca{$\tV$} of the Double Fusion Algebra
  } \\
%
\noindent
(with the constraints that their spectrum is dictated by the
modular invariant matrix $Z_{i\bi}$ as in (\ref{e:consV})
  and that \cc{$\tV_{i\bi^*;\, 1}{}^1=Z_{i\bi}$ } and                
  $\tV_{ij}^T=\tV_{i^*j^*}$.) 
\medskip

\noindent
{\bf Remark}. From a physical perspective, these defect lines or seams 
generalize cases which were known before and which were
constructed using symmetries of an underlying 
lattice model. For example, a closed non-contractible disorder 
line of the Ising model, which flips all the couplings 
along the edges that it crosses, 
is an actual realization of a certain operator $X_x$
with $x$ corresponding to a ``simple current''. 
Using the $\Bbb{Z}_2$ invariance of the model, this line may be freely 
moved around  along one of the two cycles of the torus,
in a discrete analogue of the
condition (\ref{e:seam}). This observation may offer a partial
justification to the denomination ``quantum symmetries'' that 
A. Ocneanu has given to 
objects labelled by $x$  \cite{ocneanu:1998}. Hopefully, 
the work in progress \cite{pearceetal,pearceetal2} on the actual realization
of all the operators $X_x$ in lattice RSOS models will improve 
our intuition about these operators.


\section{ RCFT on half-plane or annulus}

A consistent CFT must lead to a sensible quantum theory on any
two-dimensional manifold (world-sheet), irrespective of whether
it has or not boundaries. In the presence of boundaries, however, 
we expect constraints of a new nature to emerge, as we are probing 
different features of the theory.

\subsection{Cylinder: boundary states, Cardy condition}
Let's review the case of a cylinder with boundary conditions 
$a$ and $b$ (yet to be determined) imposed at its two ends. 
As the discussion of this case 
is  quite parallel to the one of section 3.2, we  shall be brief, see
\cite{BPPZ:1998,BPPZ:2000}.
Once again, we shall impose the consistency between two pictures:
 
\noindent\begin{minipage}[t]{0.65\textwidth}
\ca{(1)} On the one hand, the system may be described in terms of
boundary states $|a\rangle$, $|b\rangle$ in $\CH$ satisfying 
$$ (L_n-\bL_{-n}) |a\rangle =0$$ (and likewise for $|b\rangle$) and 
their generalizations to other generators $W_n$. 
One proves (again by a suitable
application of Schur's lemma) that there exists a canonical 
(``Ishibashi'') such state \cc{$|j,\za\rrangle$} 
\end{minipage} \hfill
\begin{minipage}[t]{0.33\textwidth}\vspace{0pt}
\centering \includegraphics[width=0.8\textwidth]{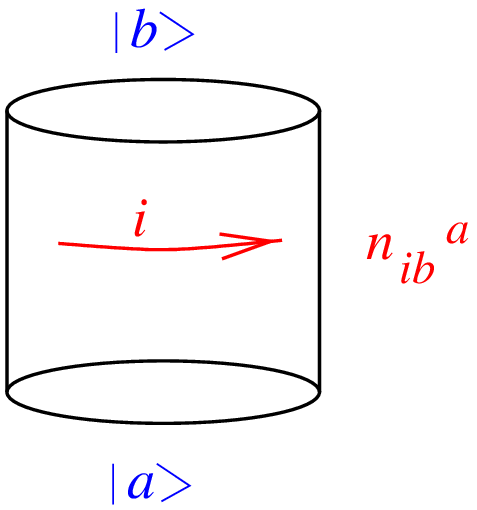}
\end{minipage}
\vskip1mm \noindent  
in each 
 $\CV_j\otimes \CV_{\bj}$ iff $j=\bj$ (up to automorphisms, see 
\cite{BPPZ:2000} and references therein)
and $Z_{jj}\ne 0$. We use 
$\za=1,\cdots,Z_{jj}$ to label a basis of such states.
The most general boundary state is thus written as a linear superposition
$$\qquad |a\rangle=\sum_{j,\za=1,\cdots, Z_{jj}} 
{\psi^{(j,\za)}_a\over \sqrt{S_{1j}}}|j,\za\rrangle\ .$$
(2) On the other hand, the Hilbert space of the theory 
in the upper half-plane supports the action of a single 
 copy of Vir or $\gA$   and thus decomposes as 
\beq
\CH_{ba}=\oplus_{i\in\CI}\ n_{ib}{}^a\ \CV_i
\label{bhilb}\eeq
with integer multiplicities $n_{ib}{}^a$.
We now compute the partition function of the system on a finite
cylinder in the two pictures: evolution between the boundary 
states $|a\rangle$ and $\langle b|$ on the one hand, or periodic
evolution in the Hilbert space $\CH_{ab}$ on the other. 
Consistency gives the {\bf Cardy equation} \cite{cardy:1989}: 
\beq
n_{ia}{}^b=\sum_{j,\za} {\psi^{(j,\za)}_a \psi^{(j,\za)\,*}_b}
{S_{ij}\over S_{1j}}. 
\label{e:cardy} 
\eeq
We again  assume $\psi$ to be a unitary matrix
(reflecting the completeness of the set of boundary states)    
and invoke (\ref{e:verl}) to conclude that
\beq
n_i n_j=\sum_k N_{ij}{}^k n_k\,,\qquad (n_i)^T=n_{i^*}\,.
\label{e:nimrepn}
\eeq
Hence the $\{n_i\}$ form a nimrep of the fusion algebra
\footnote{
 The Verlinde algebra nimreps (\ref{e:nimrepn}), 
introduced  in  \cite{dfz:1990}, were conjectured there to
provide the coefficients of  cylinder partition functions. 
The  completeness requirement
was first stressed in \cite{PSS} in a particular example.
However a key step of the attempted derivation of 
(\ref{e:nimrepn}) there (the identification of the 
boundary field
multiplicity $n_{ia}{}^b$  with that of bulk field couplings) is 
incorrect.}
  and this leads us to yet another \dots

\Problem{{ Classify the nimreps $n$ of the fusion algebra }} \\
\noindent
(with the constraint that their spectrum is dictated by the (diagonal 
part of the) modular invariant $Z_{jj}$ as in (\ref{e:cardy}), 
and subject to the symmetry  $n_{i^*}=n_i^T$.)

This problem has already been investigated for a while.
Since in our setting the spectrum is restricted 
to that dictated  by the modular invariant we automatically
discard as unphysical any would be solution of the system (\ref{e:nimrepn}) 
not consistent with modular invariance: see below in sect. 5.2
an example of such a spurious solution. On the other
hand there may be more than one solution corresponding to the same
modular invariant. Some of those may have to be discarded by further
requirements,  see the discussions below.

Before we show explicit solutions of these nimreps, 
let us complete our guided tour by combining defects and boundaries.


\subsection{ Cylinder and Defect Line}

It is very natural to combine the two situations above and 
to look at the spectrum of the theory
when both defect lines and boundaries are present. 

\noindent
\begin{minipage}[t]{0.65\textwidth}
One finds that on a cylinder with one defect line $X_x^\dagger$ and 
boundary states $a$ and $b$, a
new set  of integer multiplicities occurs in the decomposition of   
$\CH_{bx|a}=\oplus_{i\in \CI} (n_i\,\tn_x)_{b}{}^{a} \CV_i$
\beq
\tn_{ax}{}^b=\sum_{j,\Ga,\Gb}\psi_a^{(j,\Ga)}{\Psi_x^{(j,j;\Ga,\Gb)}
\over\sqrt{S_{1j} S_{1\bj}} 
}\psi_b^{(j,\Gb)*} \ .
\label{e:ntilde}
\eeq
\end{minipage} \hfill
\begin{minipage}[t]{0.30\textwidth}\vspace{0pt}
\centering\includegraphics[width=0.85\textwidth]{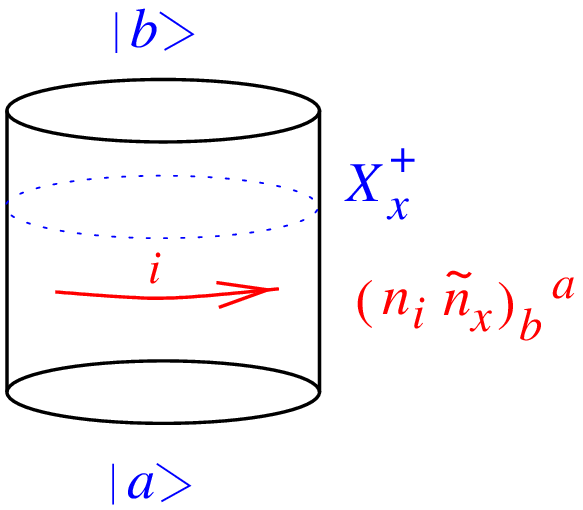}
\end{minipage}

The structure of equation (\ref{e:ntilde}) looks familiar but
note  that in contrast with the cases encountered so far, 
it doesn't express the full diagonalization of the matrices $\tn$,
at least whenever some $Z_{jj}>1$, resulting in a 
non-diagonal 
sum over $\za,\zb$. In these conditions, the $\tn$ {\bf do not}
commute a priori.  
Indeed the $\tn$ form a nimrep of a new, associative
but non-commutative, algebra, endowed with a $*$ involution 
\cc{\begin{eqnarray}
\tn_x \tn_y&=& \sum_z \tN_{xy}{}^z \tn_z\nonumber\\
\tN_{yx}{}^z&=&\sum_{j,\bj;\Ga,
\Gb,\Gc}\, \Psi_y^{(j,\bj;\Ga,\Gb)}\,
 {\Psi_x^{(j,\bj;\Gb,\Gc)}
      \over\sqrt{S_{1j} S_{1\bj}}}\,
\Psi_z^{(j,\bj;\Ga,\Gc)*} 
\label{e:tildeN}\\
\tN_x \tN_y&=& \sum_z \tN_{xy}{}^z \tN_z \ .\nonumber
\end{eqnarray}}
\noindent
Since $\tn_{ax}{}^b$ are interpreted as multiplicities  
both  (\ref{e:ntilde})
and (\ref{e:tildeN}) have to give non-negative integers.
Note that it is quite non-trivial that 
bases $\Psi$ and $\psi$ may be found such that the integrality of
the $\tn\,, \tN$ holds true.
As a consequence of (\ref{e:consV},\ref{e:tildeN}) and of
the unitarity of $\Psi$, the $\tV_{i\bi;\, x}{}^y$ taken for fixed
$i,\bi$ satisfy 
\beq
\tV_x{}^y =\sum_z \tN_{xz}{}^y \  \tV_1{}^z \ .
\label{e:tVtN}
\eeq

 Physically, the algebra with the structure constants  
 $\tN_{yx}{}^z$ in  (\ref{e:tildeN})
describes the {\it fusion of defect lines}\,;
(\ref{e:tVtN}) says that once the $\tN$ are known, insertion 
of an arbitrary number of defect lines reduces to a single one; 
and the fact that this algebra is non-commutative expresses that 
defect lines cannot be freely swapped along the cylinder. This 
non-commutativity
takes place whenever the Hilbert space (\ref{e:hilb}) contains 
multiplicities larger than 1. The simplest statistical mechanical
model in which this is expected and where it has now been 
observed on the lattice is the 3-state Potts model \cite{pearceetal2}.



\section{From Nimreps to Graphs and Cells}

\subsection{The system of nimreps}
Let us summarize where we are. We have found that 
the spectrum of a RCFT in various ``environments'' is 
described by a set of multiplicity matrices which have the 
remarkable property of forming nimreps of fusion algebras 
\cc{
\begin{eqnarray}
N_i N_j&=&\sum_k N_{ij}{}^k N_k
\nonumber\\
n_i n_j&=&\sum_k N_{ij}{}^k n_k\nonumber\\
\tV_{i_1j_1} \tV_{i_2j_2}&=&
\sum_{i_3,j_3} N_{i_1i_2}{}^{i_3} N_{j_1j_2}{}^{j_3}\
\tV_{i_3j_3} \qquad\qquad \label{e:system}\\
\tn_x \tn_y&=& \sum_z \tN_{xy}{}^z \tn_z\nonumber\\
\tN_x \tN_y&=& \sum_z \tN_{xy}{}^z \tN_z  \,.
\nonumber
\end{eqnarray}}
\noindent 
Assuming in addition that
the matrices $n_i$ and $\tn_x$ commute, the second and the
fourth of these equations can be combined into
\beq
\big(n_i \tn_x\big)\, \big(n_j \tn_y\big)=\sum_{k,z}\, N_{ij}{}^k\,
\tN_{xy}{}^z\ \big(n_k \tn_z \big) \ .
\label{double} \eeq

These various sets of matrices cannot be studied completely 
independently as they are connected by various types of relations: 
multiplicities of eigenvalues 
of $n_i$, $\tV_{ij}$ given in terms of $Z$, 
commutation relations $[n_i,\tn_x]=0$, etc. 
The system (\ref{e:system}) simplifies in the  {\it diagonal} cases
in which $Z_{ij}=\delta_{ij}$.
Then the ranges of all labels coincide and (\ref{e:system}) is solved 
with $n=\tn=\tN =N\,, \tV_{ij}=N_i N_j$.



\subsection{Graphs}

The (integer valued) matrices $n_i\,,\tV_{ij}$ 
are conveniently encoded as graphs.
Each of them may be regarded as the adjacency matrix of a graph
whose vertices are labelled by the set of indices of the matrix at
hand. As these matrices form a nim{\it rep} of an algebra and are thus
reconstructed from the generators of that  algebra, it is 
sufficient to draw the graph(s) corresponding to the generator(s).
For example, for
{$\widehat{sl}(2)$, it is sufficient to give $N_2$, $n_2$, 
and $(\tV_{21},\tV_{12})$ to reconstruct all $N_i,\ n_i$ and 
$\tV_{ij}$. 

{\it  Graphs of $n$. } Consider first the simple case of 
$\widehat{sl}(2)_k$. By (\ref{e:cardy}), $n_2$ must have eigenvalues 
$S_{2j}/S_{1j}=2\cos \pi j/(k+2)<2$, for some $j\in \CI$
hence be the adjacency matrix graph of an $A$-$D$-$E$ Dynkin diagram 
or a ``tadpole'' $A_{2n}/\Bbb{Z}_2$. Tadpoles are ruled out by the
spectral condition as their eigenvalues do not
match the $ADE$ classification of modular invariants. 
For the case  of $\widehat{sl}(3)$, enough nimreps have been
found along the years  
\cite{kostov,dfz:1990,BPPZ:2000} to match Gannon's list 
of modular invariants \cite{gannon:19xx}
and it is now believed \cite{ocneanu:2000} that 
the list is complete. See \cite{BPPZ:2000} for that list
and \cite{fhssw:00} for another systematic study 
 in the case of simple current modular invariants.

Given a nimrep $\{n_i\}$ of matrices of size $p\times p$ and 
its set of graphs, it proves useful to introduce their 
{\it associative graph algebra}: it is realized by $p\times p$ matrices 
$\hN_a=\{\hN_{ba}{}^c\}$,
$\hN_a \hN_b=\sum_c \hN_{ab}{}^c \hN_c$,
  such that  $\hN_1=I_p$, 
the identity matrix,  for some special vertex denoted 1.
The $\hN_a$ are requested to satisfy
\beq
n_i \hN_a=\sum_b n_{ia}{}^b \hN_b\ . \label{e:nhatN}
\eeq
Thus, if matrix $\hN_a$ is attached 
to vertex $a$, multiplication by $n_i$ gives the sum of its
``neighbours'' on the graph of $n_i$. For all known type I theories, 
(see end of sect. 3.1), one observes the following properties:
(i) such $\hN$ matrices may be found with non negative 
integer entries, see \cite{BE:xx,pzcell} for a (block) spectral 
decomposition of these --in general non-commutative-- matrices;
(ii) a subalgebra of this graph algebra associated with  a subset 
$T$ of the vertices gives the fusion algebra of the extended
chiral algebra $\gAe$ \cite{dfz2}.
 Thus the subset $T$ of vertices is in one-to-one 
correspondence with the set of representations of $\gAe$. 
One may further require that (iii) the integers $n_{i1}{}^a\,, a\in T$ 
describe the multiplicity of the representation $i$ in the 
representation $a$ of $\gAe$ and thus determine
the corresponding block-diagonal modular invariant, i.e.,
$n_{i1}{}^a=b^{i}_a$ in the notations  of sect. 3.1. This is  found
in many type I theories \cite{dfz2,pz:1996} but this  rules out 
some exceptional solutions of (\ref{e:nimrepn}), which are otherwise 
consistent with the spectral properties: an example is provided by 
the nimrep of $\widehat{sl}(3)_9$ called $\CE^{(12)}_3$ in the list of 
\cite{BPPZ:2000}. The properties (i-iii) of type I theories  
are incorporated
in the subfactor approach \cite{xu,BE:xx,BEK:xx}, 
where (\ref{e:nimrepn}), (\ref{e:nhatN})
are solved recursively for $n_i$ and $ \hN_a$,
 starting from the $a=b=1$ element of the matrix relation (\ref{e:nimrepn}),
and using the data $N\,, N^{\rm ext}\,, n_{i1}{}^a\,, \, a\in T$
of $\gA\,,\gAe$. The algorithm  extends to type II
with an appropriate choice of $n_{i1}{}^1$, see also section 3.5
of \cite{BPPZ:2000}. Note that (\ref{e:nhatN}) implies 
\beq
n_{ia}{}^b =\sum_c \, \hat{N}_{ac}{}^b  \,n_{i1}{}^c 
\,,
\label{e:nhatNb}
\eeq
in clear analogy with (\ref{e:tVtN}).
In particular for any $a$, $n_{ia}{}^a\ge n_{i1}{}^1$ in any type I
theory. This inequality was recently proposed in 
 \cite{fs:2001} (see also the discussion in \cite{gannon:2001}),  
as a constraint for selecting ``physical'' solutions among the nimreps
 $\{n_i\}$
and is non trivial from the point of view adopted in that paper. 
The previous discussion shows that 
in the present framework it is not very
 helpful,  as for type I theories it follows from the 
positivity of the graph algebra,  
assumed here, 
while for type II it fails,
as may be explicitly checked on examples. 
 Inserting (\ref{e:nhatNb}) into (\ref{bhilb}) provides an
  alternative decomposition of the half plane Hilbert space
 (and of the corresponding 
partition function) \cite{BPPZ:2000}
\beq
\CH_{ab}=\oplus_{c}\ \hN_{ac}{}^b \ \hat{\CV}_c \,,
\label{hilb}\eeq
in terms of the ``twisted'' (or ``solitonic'')
 representations
\beq
\hat{\CV}_c= \oplus_{i\in\CI}\ n_{i1}{}^c\ \CV_i\,, 
\label{solit}
\eeq
which are true representations of $\gAe$ for $c\in T$ in type I.
The integers  in (\ref{solit})  (or  in (\ref{e:nhatNb})), 
the intertwining matrix $n_{i1}{}^c$, can be found for all $ADE$ 
$sl(2)$ graphs in Table I of \cite{dfz:1990}.
The $\hN_c$ fusion algebra
of the  type I $sl(2)$ graphs $A\,, D_{\rm even}\,, E_6\,, E_8$ 
is at the heart of a notion of ``finite
 subgroups of the quantum group $U_q(sl(2)$'' for $q$  root of unity
in the recent paper \cite{KO}, where the decompositions (\ref{solit})
appear with a different interpretation. 
On the other hand the approach of Ocneanu to the same problem 
is based on the  fusion algebra $\tN_x$ of the $\tV$ graphs, 
to which we now turn.

\medskip  
{\it  Graphs of $\tV$. }
The graphs of $\tV$  may be less familiar, 
they were introduced by Ocneanu first in the $\widehat{sl}(2)$ case
\cite{ocneanu:1998}, see 
 \cite{ocneanu:2000,BE:xx, BEK:xx, pzcell,coq} for
 further developments or discussions.
They are depicted on the Table below for $\widehat{sl}(2)$:  the two 
graphs of $\tV_{21}$ and of $\tV_{12}$ are drawn on the same chart, 
with edges of the former in solid (red) lines, those of the
latter in broken (blue) lines. 
The matrices $\tV_{i1}$ and $\tV_{1i}$ satisfy (\ref{e:nimrepn}), 
i.e. form a nimrep of the same type as $n_i$, but their spectrum 
is in general different. In particular, the graph of $\tV_{21}$, say,
is in general made of several connected components of $ADE$ type. 
This results from (\ref{e:consV}) which tells us that 
in this  $\widehat{sl}(2)$ case
 $\tV_{21}$ has eigenvalues  $ 2\cos j\pi/h$ with multiplicity
$\sum_{\bj} Z_{j\bj}^2$. It follows from the explicit form of the
$Z$'s that 
\begin{eqnarray}
A_{n} &&  \tV_{21}\sim A_n \nonumber\\
D_{2 \ell+1} &&\tV_{21}\sim A_{4\ell-1}\nonumber\\
D_{2\ell} &&
\tV_{21}\sim   D_{2\ell} \oplus   D_{2\ell} \\
 E_6 &&  \tV_{21}\sim E_6 \oplus E_6 \nonumber\\
E_7 &&\tV_{21}\sim D_{10} \oplus E_7\nonumber\\
E_8 &&  \tV_{21}\sim E_8 \oplus E_8 \oplus  E_8 \oplus E_8\nonumber
\end{eqnarray}
where $\sim$ means the equality in a certain basis.
The same holds for $\tV_{12}$, but in a different basis, and 
it is more difficult to see how these two
graphs intertwine into the patterns of the Table.
It seems reasonable to conjecture that in general the  graphs of
$\tV_{i1}$ and of $\tV_{1i}$ always contain a connected component
isomorphic to the  graph of $n_i$ relative to the ``parent''
theory.

\bigskip
\includegraphics[width=0.95\textwidth]{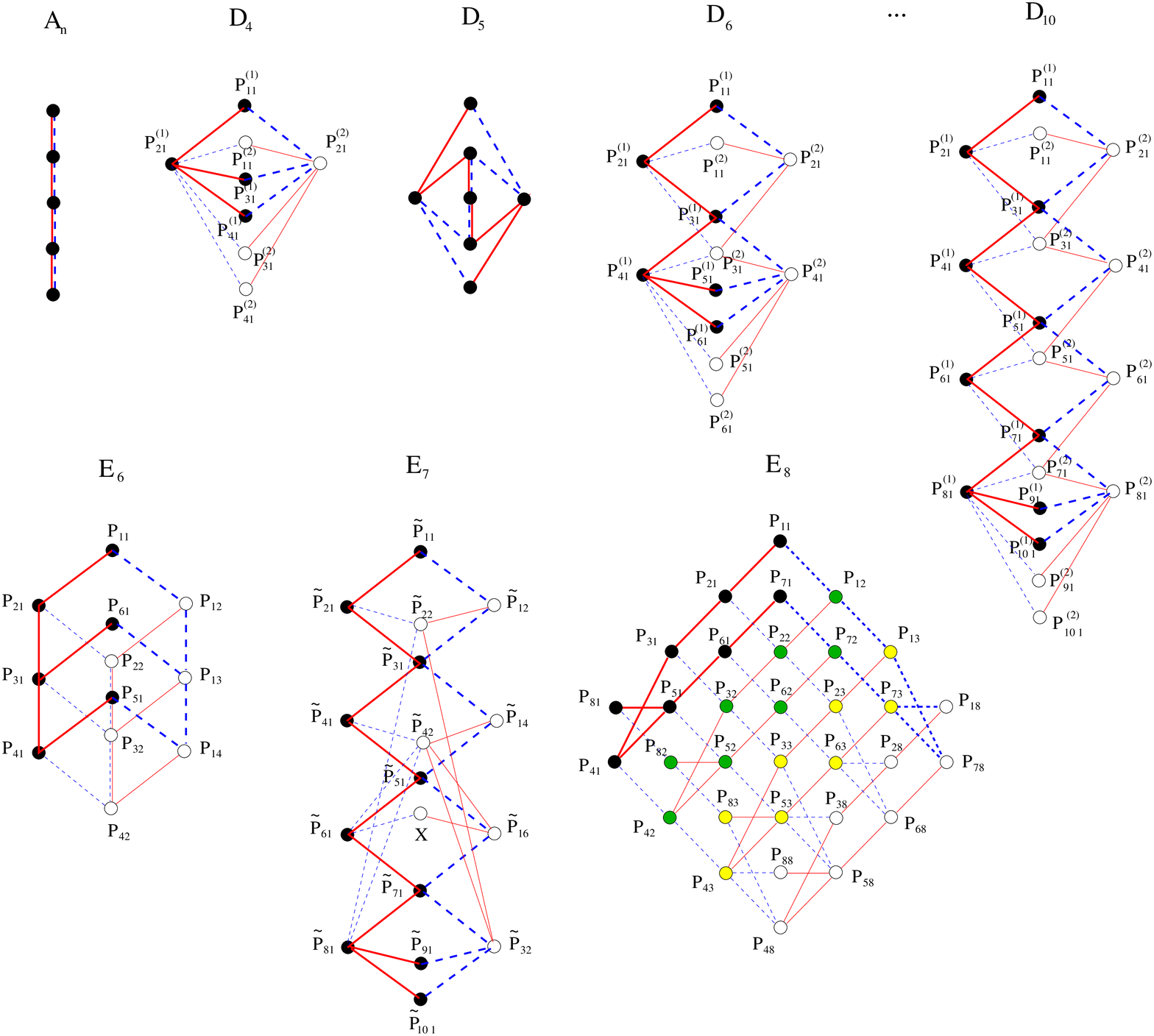}
\medskip

Note that  $\tV_{ij} \tN_x=\sum_z \tV_{ij\,x}{}^z \tN_z$
in analogy with (\ref{e:nhatN}): the $\tN$ matrices form the
graph algebra of the Ocneanu graph, and it is thus natural to 
attach $\tN_x$ to the vertex $x$ of the latter.
Moreover, one proves that there exist two special  vertices
denoted 2 and $\bar 2$ such that 
\beq
 \tN_2= \tV_{21}\qquad \tN_{\bar 2}=\tV_{12}\,.
\eeq
This follows from the observation that (\ref{e:nimrepV},\,\ref{e:tVZ}) 
gives $\sum_x (\tV_{21;1}{}^x)^2= Z_{11} +Z_{13}=1$,  as $Z_{13}=0$
for all $\widehat{sl}(2)$ modular invariants. Thus there exists a unique vertex
$x:=2$ such that  $\tV_{21;1}{}^x=1$. Then from (\ref{e:tVtN}), 
$\tV_{21}=\tN_2$ and a similar property for $\tV_{12}$. 
This discussion extends for all $\widehat{sl}(N)$ to $\tV_{f,1}$
  where $f$ is the fundamental 
representation with a single box Young tableau.


In contrast with their counterparts
$\hN_a$, the matrices  $\tN_x$ are nonnegative integer valued
 for the type I  and type II theories alike. 
 According to the previous conjecture on the existence of a connected 
 component  of Ocneanu graph isomorphic to the graph of the $n_i$, 
 in type I cases, the $\hat N$ algebra is represented isomorphically
  by a subalgebra of the Ocneanu graph algebra $\tN$, 
while in type II cases it is  the parent $\hat N$  graph algebra 
which  is a subalgebra of the parent $\tN$ algebra; 
 recall that $\tN$ is required to be positive  in general.
 Thus for the Ocneanu  graphs 
the type I versus type II distinction is less conspicuous, and
the existence in all cases of a fusion ring associated with 
each of these
graphs plays an important role in the construction of the corresponding
 quantum algebra, see below.


Instead of attaching matrix $\tN_x$ to vertex $x$ in the graph
of $\tV_{21}$ and $\tV_{12}$,  one may also attach to it
the matrix $\tV_1{}^x$. Upon multiplication 
by $N_2$, the usual fusion matrix, acting on the first label $i$ of
$\tV_{ij}$, one gets, by virtue of
(\ref{e:nimrepV}), $ \sum_{i'}N_{2i}{}^{i'} \tV_{i'j;1}{}^x=
\sum_y\tV_{21;y}{}^x \tV_{ij;1}{}^y$. Likewise, action of $N_2$ on the 
second label  is  represented by $\tV_{12}$. 
The annotations on the figures of the Table encode some 
information on this representation: the matrices $V_{ij;\, 1}{}^x$, 
forming the ``toric structure'' of  Ocneanu \cite{ocneanu:2000},
may be expressed in terms of quadratic combinations of the $n$
matrices, $\tV_{ij} =\sum n_i \times  n_j$, 
where we are deliberately vague about the
summation, the correspondence between the indices of $\tV$ and those
of the $n$'s etc. See section 7 and Appendix B of \cite{pzcell}
for a detailed discussion. More precisely, 
one must distinguish between the type I theories 
for which this holds true, and the type II ones,
here the  cases $D_{2\ell+1}\,, E_7$, for which 
this is verified in terms of the $n$ matrices of the parent type I
theory ($A_{4 \ell-1}$ and $D_{10}$ respectively). 
The above factorization of the
 matrices $\tV_1{}^x$ is paralleled by a factorisation of the multiplicities
$\tn_x$ and $\tN$, which are expressed in terms of the $n_i\,,\hat N$ data of
either the graph $G$ or its parent graph.



\subsection{Cells}
The previous multiplicities (or graphs) specify the spectrum of the RCFT.
But as is familiar, beyond the spectrum, we need additional 
information on the  ``Operator Product Algebra'' to determine fully
the theory and be able to compute all correlation functions. This 
requires to attach other quantities, which following Ocneanu 
we call {\it cells}, to these graphs.

Together, as we shall see in the next section, these  data enable one to 
construct a new quantum algebra \cc{$\CA$} (and its dual $\hat{\CA}$).
Its description is conveniently achieved by first constructing a 
simplicial 3-complex made with the elements of the next figure
\cite{ocneanu:1993,bsz}.


The 1-simplices (edges) carry labels of the nature
discussed in the beginning of these notes: $i$ is a representation label,
$a$ a boundary state, $x$ a defect. Each 2-simplex (triangle) 
comes with a multiplicity label which takes
$N_{ij}{}^ k\,,\, n_{ja}{}^c\,,\,
\tn_{ax}{}^c\,, \, \tN_{xy}{}^z\,$ values, respectively.
Finally to each of the 3-simplices
(which we have not decorated with their indices, for more clarity), 
one must assign a complex number called a cell 
(the value of a 3-cochain). 
\vskip-3mm \noindent 
For example  
\includegraphics[width=0.10\textwidth]{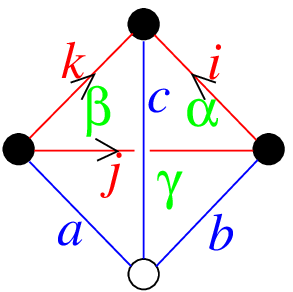}
$ = 
\Fo_{b k}\left[ 
{i\atop c}{j\atop a}
\right]_{\Ga\,\Gc}^{\Gb\, t}$, $t=1,\cdots, N_{ij}{}^k$, 
$\alpha=1,\cdots, n_{ib}{}^c\,,$ 
etc.
\vskip5mm 
\medskip\includegraphics[width=0.8\textwidth]{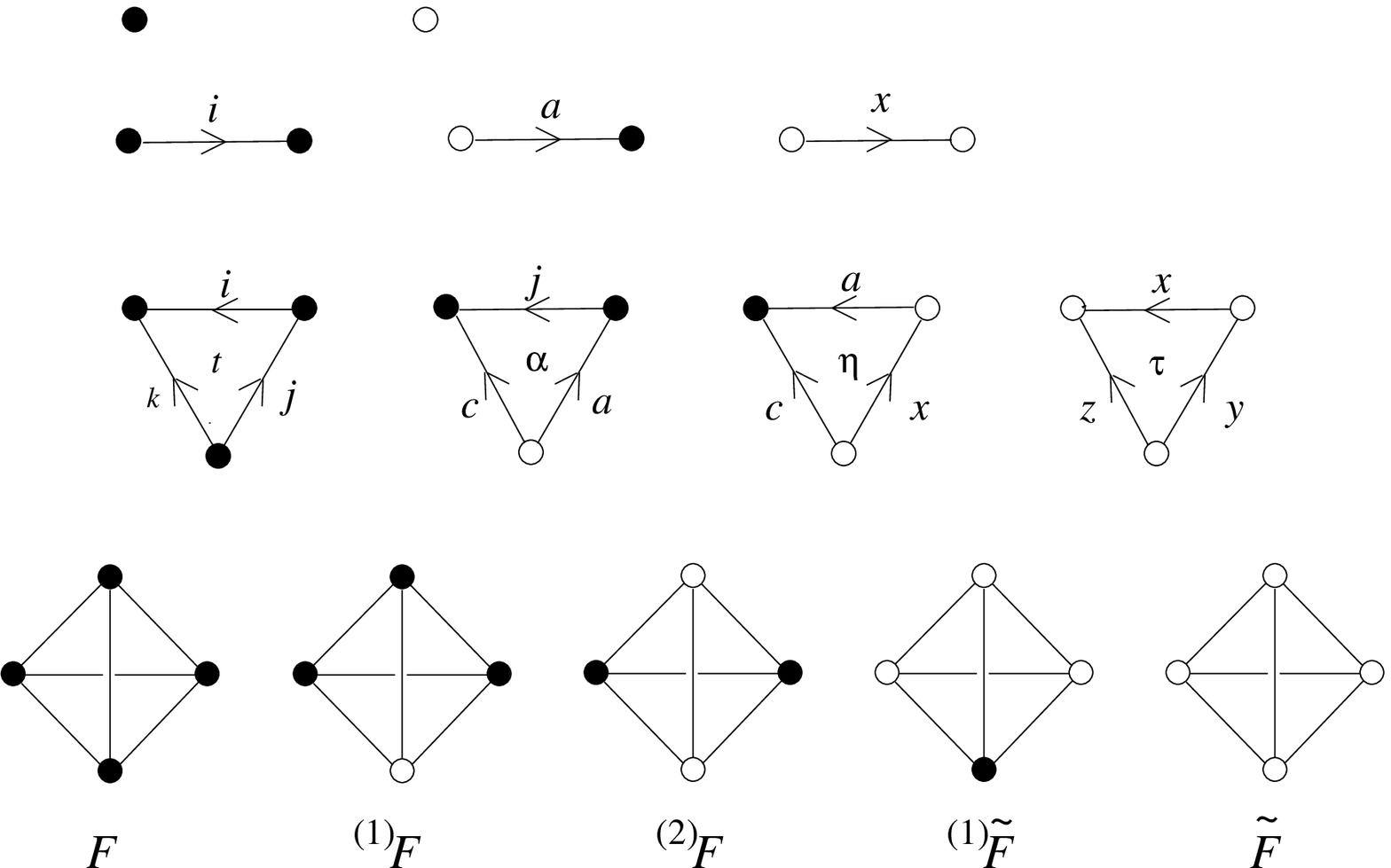}

\bigskip


In the same way as in section 2 where $F$ was expressing a change of
basis in the fusion of CVO's, all the cells
should be regarded as invertible matrices. The fact that 
$F\,,\,\Fo\,,\, \tFo$\,,\, $\tilde F$
 are square matrices is precisely one of the conditions of the big 
system (\ref{e:system}). For example 
$ \Fo_{b k}\left[ {i\atop c}{j\atop a}
\right]_{\Ga\,\Gc}^{\Gb\, t}$ has $ (b,\za,\zg)$  as a ``row  index''
and $(k,\zb,t)$ as a ``column index'' and
$$\sum_{b,\za,\zg}1=\sum_b n_{ja}{}^b n_{ib}{}^c= \sum_k N_{ij}{}^k
n_{ka}{}^c= \sum_{k,\zb,t}1\ .$$
Similarly  the spectral decompositions  of
the mutually commuting $n_i$ and $ \tn_x$ imply the relation
 $\sum_x \tn_{ax}{}^{a'} \tn_{b'x^*}{}^b=
\sum_i n_{ia}{}^b n_{i^* b'}{}^{a'}$ which ensures the invertibility 
of $\Ft$.

\noindent
\begin{minipage}[t]{0.75\textwidth}
This system of cells may be chosen  to satisfy unitarity constraints. 
More crucially, it  must satisfy the ``Big Pentagon equation''
(a name adopted in \cite{bsz}), namely a  set of 
$6$ quintic identities of the form 
\bea
 F F F &=&  F F \,,\qquad F\, \Fo\, \Fo =
\Fo\, \Fo\,,\\ 
\Fo\, \Ft\, \Ft &=&
\Fo\, \Ft \hbox{,\quad etc} \cdots 
\eea
and their dual counterparts. It may be pictorially inter-
\end{minipage} \hfill
\begin{minipage}[t]{0.20\textwidth}\vspace{0pt}
\centering\includegraphics[width=0.95\textwidth]{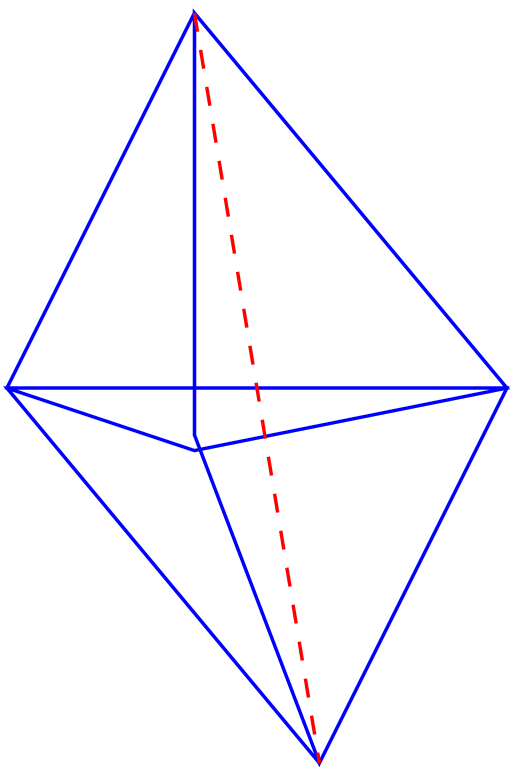}
\end{minipage} 
\vskip1mm \noindent
preted as expressing 
in\-dependence of the cochain with respect to the splitting  
of a double  tetrahedron into two or three tetrahedra: see figure.
\omit{ \end{minipage} \hfill
\begin{minipage}[t]{0.20\textwidth}\vspace{0pt}
\centering\includegraphics[width=0.95\textwidth]{pentagon.eps}
\end{minipage} 
\smallskip}
There, all assignments of the two colours to the vertices are allowed, 
but the number of identities is reduced by the unitarity of the cells.

Some of these cells and their pentagon equations
have a direct physical interpretation, notably
the second pentagon above    expresses the associativity 
of the boundary fields  product, with the cells $\Fo$ 
serving as OPE coefficients \cite{runkel:1999,BPPZ:2000}.
 See \cite{pzcell} for their lattice model interpretation.
There is an increasing evidence that the consistency 
of the RCFT requires the existence of the whole system, which leads us
to a formulation of the \dots

{\bf Refined problem :}
{\sl Find a system of nimreps (\ref{e:system}) leading to cells 
 consistent with the pentagon identities.}

In fact very little is known about this problem. 
 In the diagonal theories all the cells coincide, in an
 appropriate gauge, with the
fusing matrices $F$, explicitly  known in the $\widehat{sl}(2)$ cases.
Runkel has computed the $\Fo$ cells in the $\widehat{sl}(2)$ $D$ cases
\cite{runkel:2000} and some of the matrix elements 
 of $\Fo$ in the exceptional cases
have been worked out earlier in the context of their lattice interpretation.
In the  simpler $\widehat{sl}(2)$  $D_{\rm odd}$ case 
one has $\tN=N\,, \tn=n$, and  accordingly
$\tFo=\Fo\,, \tF=F$ 
while $\Ft$ is expressed bilinearly  in terms of $\Fo$.

Since in the type I cases the $\tN$ algebra contains a subalgebra
isomorphic to the graph algebra $\hat N$ one can 
restrict in a first step all labels in the 
 two  pentagon identities for the dual $3j$- and $6j$-symbols
   $\tFo$ and $\tF$ to  a subset identified with 
the boundary set $\{x=a\}$. Then these two relations coincide if we
identify these particular matrix elements of the two
sets of   cells $\tFo$ and $\tF$. 
In this way one introduces $6j$-symbols 
$\hat{ F}_{b d}\left[ {a\atop c}{e\atop f} \right]$
with all six labels given by boundary indices; the triangles of 
these tetrahedra are determined by the
graph fusion algebra multiplicities $\hat{N}_{ab}{}^c$.
\footnote{These $\hat F$, ``hidden''
in  the  set of dual $3j$- and $6j$-symbols of the Ocneanu algebra,
 are to be compared with the $6j$-symbols of the ``boundary category'' 
in   \cite{fs:2001}.}

We recall from sect. 5.2 that a subset $T$ of the boundary labels  can be 
identified with the representations of the extended fusion algebra.
 Accordingly $\hat{ F}$, with all $6$ labels further
restricted to $T$, can be identified with the $6j$-symbols
(the fusing matrices) $ F^{\rm ext}$ 
of chiral vertex operators  of the extended theory.

The matrices $\hat N_a$ enter as  building blocks in the expressions
for the type I multiplicities $\tn$ and $\tN$  \cite{pzcell}, 
and it is natural to  expect
 that this property can be ``lifted'' to the dual cells,
$\tFo\,, \tF\,,$ expressing them in terms of some $\hat F$. 
One can further  speculate  that the $\hat F$ themselves may be
found in terms of the 
$3j$-symbols,  extending to the cells
the relation (\ref{e:nhatNb}) and the algorithm in \cite{xu} 
 for solving  (\ref{e:nimrepn}), (\ref{e:nhatN}).



\section{The Double Triangle Algebra and (B)CFT}
In this last section we sketch how knowledge of the system (\ref{e:system})
and of the cells enables one to construct a finite dimensional 
quantum algebra with peculiar properties.

\noindent
\begin{minipage}[t]{0.68\textwidth}
As a first step we construct vector spaces \cc{$V^j$} of dimension 
{$m_j=\sum_{a,c}\, n_{ja}{}^c$} with an orthogonal basis
{$|e^{j,\beta}_{ca}\rangle $}, $\quad\beta=1,\cdots, n_{ja}{}^c$.
They correspond to the second type of triangles above, depicted dually
by 
\end{minipage} \hfill
\begin{minipage}[t]{0.30\textwidth}\vspace{0pt}
\centering\includegraphics[width=0.95\textwidth]{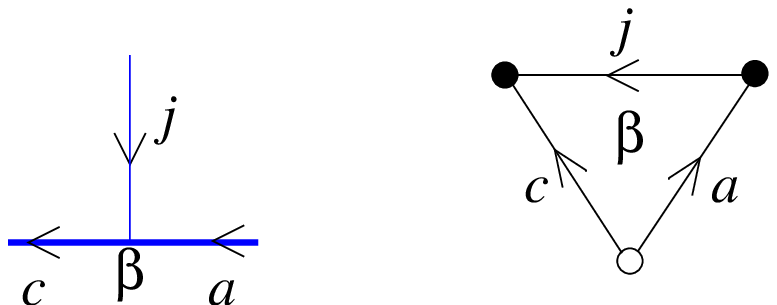}
\end{minipage} \hfill
\vskip1mm \noindent  
\noindent  generalized vertices: see figure. 
The precise normalisation of these basis 

\noindent 
\begin{minipage}[t]{0.68\textwidth}
vectors which involves
quantum dimensions and components of the Perron-Frobenius vector
  $\psi^{(1)}$ is  of no importance for the present qualitative
discussion. Then one considers the matrix 
algebra $\CA=
\oplus_{j\in \cal I}\, End\, V^j\cong
\oplus_{j\in \cal I}\,$Mat$_{m_j}\,,$ where each block 
is  generated by the ``double triangles''
$\ e_{j;\Gb,\Gb'}^{(ca)\,,(c'a')} = $
 \end{minipage} \hfill
\begin{minipage}[t]{0.30\textwidth}\vspace{0pt}
\centering\includegraphics[width=0.95\textwidth]{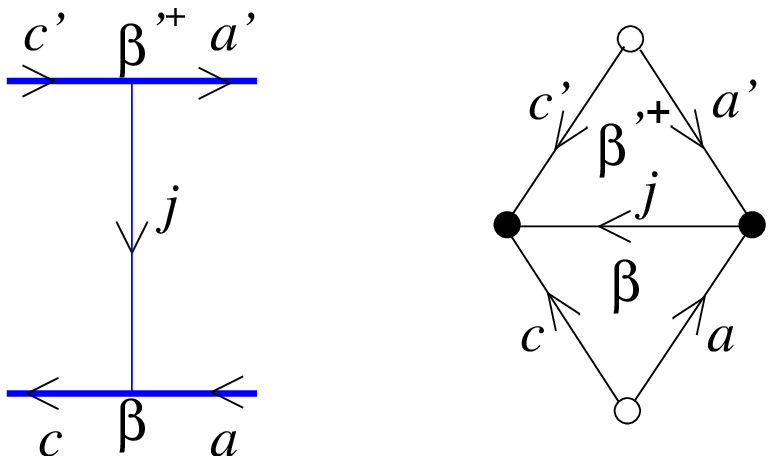}
\end{minipage}
$ \hbox{const.}
\ | e^{j\,,\Gb}_{ca}\rangle \langle e^{j\,,\Gb'}_{c'a'}| $, 
i.e., states in $V^j \otimes V^{j^*}$. The  unit 
for the matrix product is
given by $1=\sum_{j,a,c,\zb}  e_{j;\Gb,\Gb}^{(ca)\,,(ca)}$.
This algebra admits a coassociative coproduct defined 
in terms of the $3j$-symbols $\Fo$, 
a $*$- operation, counit and antipode;
the product and the coproduct can be depicted as on the picture below.

\medskip
\centerline
{\includegraphics[width=0.60\textwidth]{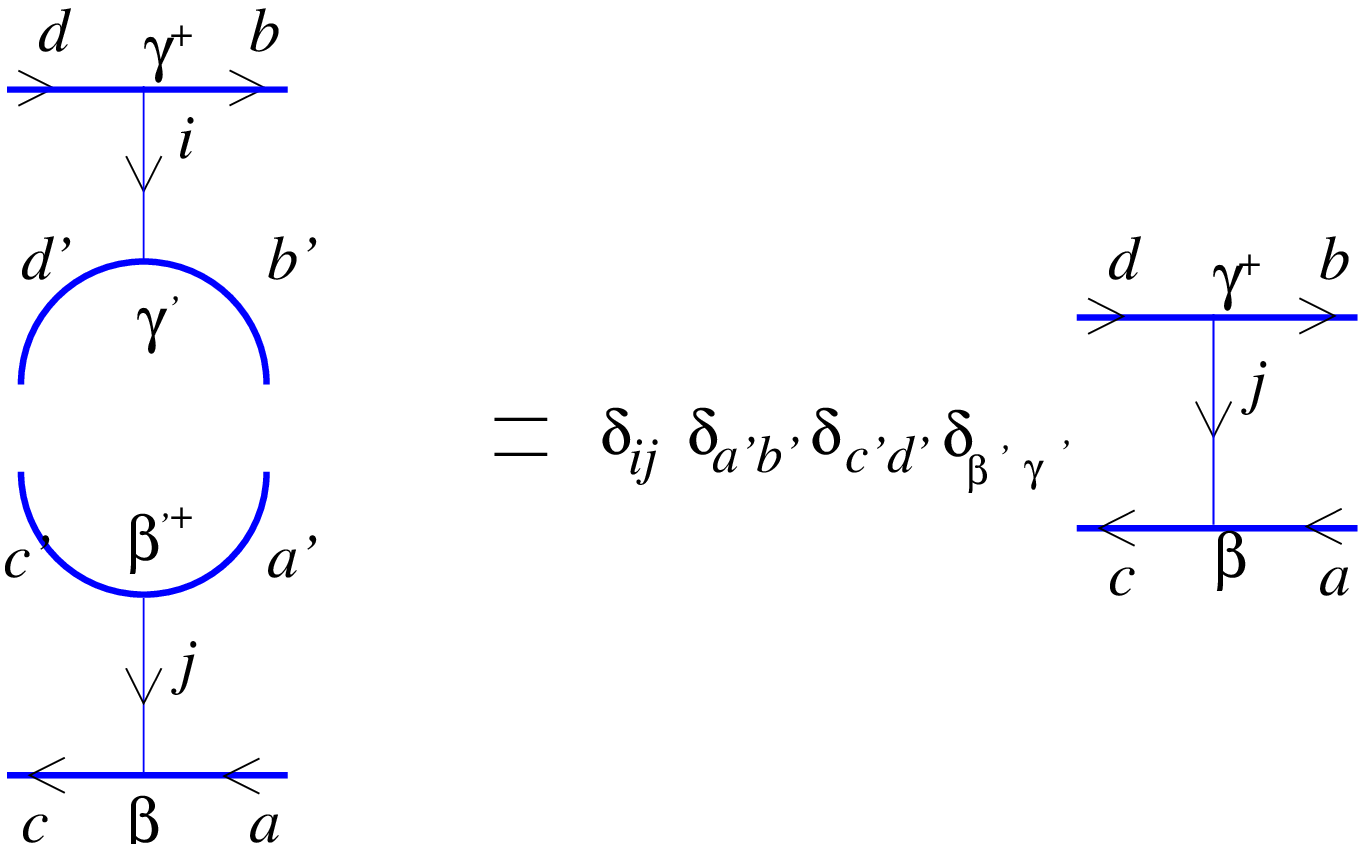}}

\includegraphics[width=0.95\textwidth]{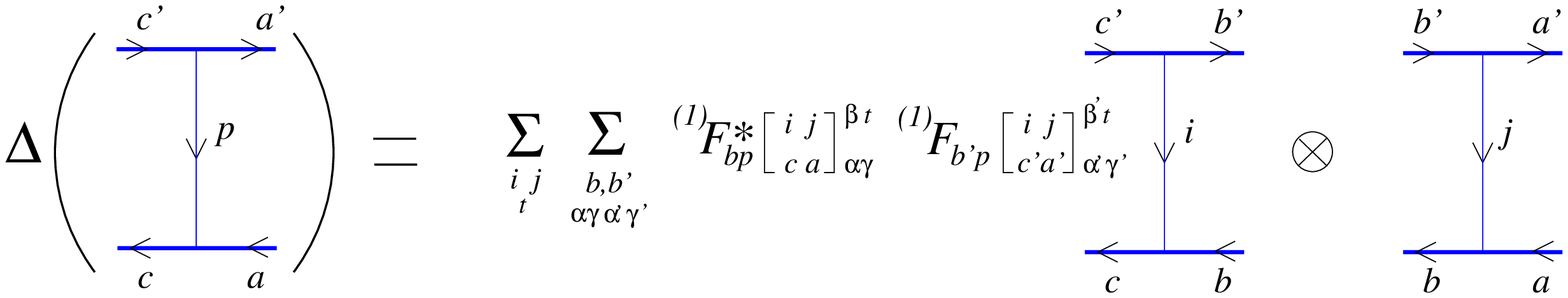}

\smallskip

These operations satisfy a slightly weakened version of the
 axioms of Hopf algebras, defining (finite dimensional)
 {\it weak $C^*$- Hopf algebras} \cite{bsz}: 
in particular  the coproduct does not preserve the identity,
$\Delta(1)\ne 1\otimes 1$,
and two of the axioms on the counit and the antipode are also relaxed.
The space of linear functionals $\hat{\CA}$ 
on $\cal A$ is given the structure of a WHA by
transposing all operations, the product, the coproduct, etc.,  
with respect to a canonical pairing $\langle\,,\,\rangle$ between 
$\cal A$ and  $\hat{\CA}$, defined with the 
help of the cells $\Ft$. In particular the dual $3j$-symbols
$\tFo$ appear in the  coproduct in $\hat{\CA}$.
The dual algebra is a matrix algebra $\hat{\CA}=\oplus_{x}\,$
Mat$_{\tilde{m}_x}\,,$ with $\tilde{m}_x=\sum_{a,b}\, \tn_{ax}{}^b$.
 As we have seen,  $n_i$ and $\tn_x$ are
(block) diagonalized in the same unitary basis, 
which ensures the equality of
dimensions of the two (finite dimensional) algebras, $\sum_i\, (m_i)^2=
\sum_x\, (\tilde{m}_x)^2 $. 
Identifying  $\cal A$ and  $\hat{\CA}$, the cells $\Ft$ are also seen 
as giving an invertible transformation relating two sets of
double triangles, i.e.,  two bases in the algebra and one can define
a scalar product consistent with the pairing
$\langle\,,\,\rangle$. Then the unitarity of
the $3j$-symbols $\Fo\,,\, \tFo$ implies that there are elements
$\hat{e}_i\,, \,  \hat{e}_j \in \CA$ and $\hat{E}_x\,, \,
\hat{E}_y \in \hat{\CA}$,  s.t. for a double triangle $e_k=e_{k;
\alpha, \alpha}^{(ac),(ac)}$ or a dual double triangle
  $E_z=E_{z;\eta, \eta}^{(ac),(ac)}$, one has
\cc{\begin{eqnarray}
 \langle \hat{e}_i \otimes  \hat{e}_j\,,\, \triangle(e_k)\rangle &=&
\sum_p\, N_{ij}{}^p\   \langle  \hat{e}_p\,, \, e_k\rangle
 = N_{ij}{}^k\\
\langle \hat{E}_x \otimes  \hat{E}_y\,,\,
\hat{\triangle}(E_z)\rangle &=& \sum_w\, \tN_{xy}{}^w\ 
\langle  \hat{E}_w\,,\,
E_z\rangle=\tN_{xy}{}^z\,.
\label{repr}
\end{eqnarray}}
In this way the two sets of integers  $N$ and $\tN$ determine 
fusion rings in each algebra of the dual pair.
In type I cases we can restrict the second equality to the subalgebras 
$\hat N$ and $N^{\rm ext}$, 
selecting accordingly subalgebras of $\hat{\CA}$. Note that (\ref{double})
plays an 
analogous r\^ole for the {\it double} $\CD(\CA)$ of the WHA $\CA$.

These pairs of WHA can be given a field theoretic interpretation.
The nimreps $n_{ia}{}^b$  provide the multiplicities of Cardy
boundary fields and at the same time count the triangles,
i.e., the states of the representation spaces $V^i$ with fixed $a,b$. 
One can then combine
the CVO $\phi_{ij}^k(z)$ with intertwiners $V^j\to V^k$ to define
fields covariant with respect to the WHA algebra $\cal A$
\cite{pzcell};  for real arguments this gives a precise operator meaning 
to the boundary fields. Furthermore the physical 
(half-plane) bulk fields of the BCFT 
are defined as  chiral compositions of the generalized CVO. 
The new constants in this construction, 
the {\it bulk-boundary } or {\it reflection} 
coefficients $ R_{a}^{(j,\bj)}(p)$,
are  defined  for $n_{pa}{}^a N_{j\bj}{}^p \not =0$.
\footnote{
Taking  different initial
and final boundary labels in these compositions 
(and accordingly bulk-boundary
coupling constants depending on  pairs 
of such labels), 
leads to more general  bulk fields, which produce
for small $z-\bar{z}$   boundary-changing fields.
It remains to check the consistency and  relevance 
of this generalization.}
They are subject of equations
 \cite{carlew}, which in particular determine them    explicitly
for scalar fields  in terms of $F$ and $\Fo$, 
see \cite{pzcell} for a general formula. All correlators 
are reduced to explicit linear combinations of conventional
chiral blocks.
 
In precisely the same way one can define in the type I cases generalized
CVO for the extended theory, which are covariant under a subalgebra of 
$\hat \cal A$, spanned by the dual double triangles with $x=a\,, a\in T$.
 These are ``diagonal'' covariant CVO, defined 
using the dual cells $\tF$ or $\tFo$, restricted to labels in 
$T$, i.e., the $6j$-symbols $F^{\rm ext}$ introduced in the previous section.

The algebra  $\cal A$ possesses an $\CR$ matrix, i.e.,
has the structure of  quasi-triangular WHA. 
Accordingly one introduces braiding matrices 
for the generalized CVO; 
they are diagonalized by the $3j$-symbols $\Fo$, while
the eigenvalues are phases determined by the scaling dimensions of
the CVO. These braiding matrices were shown also to reproduce
the Boltzmann weights \cite{ocneanu:2000, pzcell}
 of associated lattice models \cite{pasquier}.
In general there is no natural way to turn 
$\hat{\cal A}$ into a quasi-triangular 
algebra even in the commutative $\tN$ cases. 
Yet for type I cases there  are  matrices $B^{\rm ext}(\pm)$,
diagonalized by $F^{\rm ext}$, which reproduce
 the braiding matrices of the CVO
(or of the diagonal generalized CVO) of the extended theory.  

Last but not least,
the (commutative) duals of the $\hat{N}$ and $\tN$ algebras,
i.e., the {\it Pasquier algebra}   \cite{pasqu}   and its generalization,
have been shown to be highly relevant in the 
determination of the OPE coefficients of the physical bulk 
fields.
 The structure constants 
$M_{(i,\alpha)(j,\beta)}{}^{(k,\gamma)}=\sum_a \psi_a^{(i,\alpha)}\,
\psi_a^{(j,\beta)}\psi_a^{(k,\gamma)*}/\psi_a^{1}$ of the Pasquier algebra
yield the  OPE coefficients of scalar bulk fields 
normalised appropriately, \cite{pz:1995,pz:1996}, 
while their analogues constructed out of
$\Psi_x^{(j,\bj;\alpha, \alpha)}$ provide the same  information for 
the modulus square of the OPE coefficients for
 all bulk fields of the theory,
 $\Phi_{(j,\bar{j};\alpha)}(z, \bar{z})$, $ \alpha=1, 2\dots Z_{j\bj}$.
The latter formula is truly nontrivial for the exceptional type II
cases.
These universal formulae arise from  duality 
properties of  bulk fields  near boundaries
 \cite{carlew,PSS,RS,BPPZ:2000}
\footnote{
 Rediscovered in the boundary CFT \cite{PSS}, 
as resulting from the sewing equations of  \cite{carlew},
 the  algebra introduced in 1987 by Pasquier was called 
``classifying algebra'' in \cite{fs:97},
 where  its importance  in this context  was  particularly stressed.} 
or defect lines \cite{pzcell}. 

In conclusion, despite  its weakened axioms (as compared to the related
 quantum groups, say), this quantum algebra has many appealing features:
\begin{description}
\item[-] its construction relies on combinatorial data 
(multiplicities) that encode the spectrum of the RCFT for
various boundary conditions; 
\item[-] it  is consistent with the extended chiral algebra symmetry of type
  I cases;
\item[-] 
its $6j$-, resp. $3j$- symbols $F\,,\,$   $F^{\rm
  ext}\,,\,$ and $\Fo$ are intimately connected with the OPA of the RCFT; 
\item[-] it incorporates in a natural way the truncation of representations
  inherent to RCFT;
\item[-] it enables one to construct generalized chiral vertex operators, 
giving precise operator meaning to the boundary 
and the physical (half-plane) bulk fields;
\item[-] 
its  ``weakness'' is justified by the necessity of
describing the boundary field degrees of freedom, e.g., the nontrivial
``identity'' space $V^1$ of dimension 
$|V^1|= tr(n_1)$ determines the  set of vacuum states
$|0\rangle \otimes |e^1_{aa}\rangle$  in the 
correlators of boundary  and half-plane bulk fields;
\item[-] it determines the basic structure constants of the various
operator algebras for any diagonal or non-diagonal RCFT. 
\end{description}
{This qualifies $\CA$ as the natural quantum algebra of the RCFT!}



\section{Perspectives}

There are various perspectives for further developments. 

First the  $\widehat{sl}(2)$ examples have to be completed.
Although a lot of data are already available for 
 these simplest non-trivial examples of the Ocneanu quantum symmetry,
the  elaboration of all the structures involved
in the construction of the non-diagonal WHA $\CA$  remains to be done.
In general, the empirical observations or conjectures of sect. 5 
and 6 have to be established on a firmer ground.

The discussion so far was mainly aimed at the description of
the integrable WZW or the related minimal models.  
Other examples of finite dimensional WHA  arise
from the $c=1$ $\Gamma$- orbifold CFT, where $\Gamma$ is 
a subgroup of SU(2), and their generalizations to other groups
\cite{DVVV}.
The  starting point  is the complete set of  boundary states, 
described by the  representations of the quantum  double $\CD(\Gamma)$
of the finite group $\Gamma$. These are ``Cardy type'' solutions 
since there
 exists  a  Verlinde formula with  a  symmetric unitary  $S$ matrix;
restricted to the subset of untwisted  representations,
it gives simply  the representation ring of $\Gamma$
(the graph algebra of the affine Dynkin diagrams in the SU(2) case
and their generalizations).
 The cylinder partition functions can be represented either
 as finite sums in the rational characters with the Verlinde
fusion multiplicities, or as  infinite sums
 of Virasoro 
(or more generally $W$-) algebra characters,
 parametrised by the highest weights  of the finite dimensional
 irreducible representations of SU(2)  (or its  generalization),
 see, e.g., \cite{talbot} for the formulae restricted to the untwisted
 sector.
These decompositions are analogous to (\ref{hilb}) and
(\ref{bhilb}), respectively.  
The coefficients in the second sum
are  ``affine''  analogs of the nimreps $n_{ia}{}^b$;
in the untwisted sector $\{n_{i1}{}^a\}$ reproduce
the multiplicities of the representations of $\Gamma$ in the
decomposition of the finite dimensional irreps $\{i\}$.

Another possible direction  is the generalization of these
ideas to     non-rational CFT with a continuous spectrum. 

\medskip
 \Ack
\medskip

We thank  R. Coquereaux,  Ch. Mercat and C. Schweigert
for useful remarks on the first draft of this paper.

\vskip 2pc

\bibliographystyle{plain}

{\noindent \em Valentina Petkova, School of Computing and Mathematics, 
  University of Northumbria, NE1 8ST Newcastle upon Tyne, UK,} and \\
{\em Institute for Nuclear Research
and Nuclear Energy, 72 Tzarigrasko Chaussee, 1784 Sofia, Bulgaria,
 e-mail: {\tt petkova@inrne.bas.bg }}

 and

{\noindent \em Jean-Bernard Zuber,
Service de Physique Th\'eorique, 
CEA Saclay, 91191 Gif-sur-Yvette Cedex, France,
 e-mail: {\tt zuber@spht.saclay.cea.fr }
    }

\end{document}